\documentclass[pra,a4paper,twocolumn,showpacs,superscriptaddress]{revtex4}
\usepackage{amsmath}
\usepackage{amsfonts}
\usepackage{graphicx}
\usepackage{longtable}
\newcommand{\be}{\begin{equation}}
\newcommand{\ee}{\end{equation}}

\newcommand{\ba}[1]{\left(\begin{array}{#1}}
\newcommand{\ea}{\end{array}\right)}
\begin{document}
\title{\textbf{Effectiveness of Depolarizing noise in causing sudden death of entanglement } }
\author{K.O. Yashodamma }
\affiliation{Department of Physics, Kuvempu University, 
Shankaraghatta, Shimoga-577 451, India} 
\author{P.J. Geetha }
\affiliation{Department of Physics, Kuvempu University, 
Shankaraghatta, Shimoga-577 451, India}
\author{Sudha}
\email{arss@rediffmail.com}
\affiliation{Department of Physics, Kuvempu University, 
Shankaraghatta, Shimoga-577 451, India}
\affiliation{Inspire Institute Inc., Alexandria, Virginia, 22303, USA.}

\begin{abstract} 
Continuing on the recent observation that sudden death of entanglement can occur even when a single qubit of a 2-qubit state is exposed to noisy environment(Results in Physics,3,41--45 (2013)), 
we examine the local effects of several noises on bipartite qubit-qutrit and qutrit-qutrit systems. In order to rule out any initial interactions with environment, we consider maximally entangled pure states of qubit-qutrit and qutrit-qutrit systems for our analysis. We show that depolarizing and generalized amplitude damping noise can cause sudden death of entanglement in these states even when they act only on one part of the system. We also show that sudden death of entanglement occurs much faster under the action of depolarizing noise when compared to that due to generalized amplitude damping.  This result strengthens the observation (Results in Physics,3,41--45 (2013)) that depolarizing noise is more effective than  other noise models in causing sudden death of entanglement.      
\end{abstract}
\date{\today}
\maketitle
\section{Introduction} 
The pure multipartite maximally entangled quantum states are of wide scope in various fields of quantum computation, quantum information and quantum communication~\cite{nc} as these states produce  protocols with high precision. But due to an inevitable coupling of quantum systems with their surrounding environment, the states lose their coherence as well as entanglement. This irreversible interaction of quantum states with surrounding noisy environments is called decoherence~\cite{dec} and in addition to making them more mixed, it causes degradation of the initial entanglement~\cite{dec}. The study of dynamics of the quantum states, pure as well as mixed, exposed to noisy environment is an important area of study in order to devise ways such that the system can be protected from the detrimental effects of the environment.   

Depending upon the environment surrounding the quantum state there may either be an asymptotic decay of entanglement or the entanglement may vanish in finite time. The phenomenon of finite time disentanglement or Entanglement Sudden Death (ESD) has been given enough importance \cite{esd,esd1,esd1a,esd2,esd3,esd4,n10,esd5,esd6,esd7,esd8,esd9,esd9a,esd10,esd11,esd12,esd13,esd14,esd15,esd16,esd17,esd18,esd19} and considerable amount of work is also going on to avoid ESD \cite{esdpr1,esdpr2,esdpr3,esdpr4,esdpr5,esdpr6,esdpr7}.   

It has been shown in Ref. \cite{esd1} that the entanglement loss occurs in finite time under the action of pure vaccum noise in  bipartite states of qubits.  Such a finite time disentanglement is shown to be due to non additivity of  decay rates \cite{esd2} of the two parts individually exposed to either same or different noisy environments. In a recent work \cite{esd19} it is shown that, the single noise acting on a single qubit is also sufficient to cause ESD in  $2$-qubit states. It was also shown that when acted on one of the qubits of a pure $2$-qubit state, both amplitude and phase noise can only cause asymptotic decay of entanglement while depolarizing noise causes sudden death of entanglement thus establishing the effectiveness of depolarizing noise in causing ESD than amplitude noise and dephasing noise. In this article, we continue on the work in Ref.~\cite{esd19} and examine the effect of several noisy environments {\emph {on a part of}} bipartite qubit-qutrit and qutrit-qutrit systems. The noise models that we consider include amplitude damping, dephasing, depolarizing and generalized amplitude damping (GAD). We show that the local action of depolarizing noise and generalized amplitude damping noise  can cause sudden death of entanglement in pure $2\times 2$ and $2\times 3$ dimensional systems even when any one of these noises act on a part of the system.  As in the case of $2$-qubit pure states, the amplitude and dephasing noises are shown to cause asymptotic decay of entanglement only whereas both depolarizing and Generalized amplitude damping noises induce sudden death of entanglement. We compare the time scales of complete disentanglement in these two cases and show that depolarizing noise kills the entanglement in a shorter time than due to Generalized amplitude damping.    

The article is divided into four parts.  Section 1  contains  introductory remarks and motivation behind this work. In Section 2, the mode of decay of entanglement of a qubit-qutrit system in a pure state, with only the qubit  being exposed to different noisy environments is studied. Section~3 contains the similar analyis for qutrit-qutrit systems.  Section 4  provides a concise summary of the results  

\section{Action of noise on pure qubit-qutrit states}
It is well known that an arbitrary pure qubit-qutrit state
\be 
\label{ye}
\vert \psi_{qq'} \rangle =\sum_{i,j} {a_{ij} \vert i,j \rangle }
\ee
where  $i=0,\,1$ and $j=0,\,1,\,2$ and the coefficients $a_{ij}$ are in general complex with $\sum_{i,j} {\vert a_{ij} \vert^2} = 1$.   
On Schmidt decomposition $({\ref{ye}})$ takes the form  
\be
\label{yc}
\vert \psi_{qq'} \rangle \equiv a \vert 0_1 0_2' \rangle + \sqrt{ 1- a^2 } \vert 1_1 1'_2 \rangle 
\ee
with $\alpha$ being a real number $0\leq a \leq 1$; $\vert 0_1\rangle,\,\vert 1_1\rangle$ are the Schmidt bases in the $2$-dimensional space of qubits and 
$\vert 0'_2\rangle,\,\vert 1'_2\rangle,\,\vert 2'_2\rangle$ are the Schmidt bases in the $3$-dimensional qutrit space.  From Schmidt decomposed form in Eq. (\ref{yc}) a maximally entangled pure qubit-qutrit state can be readily seen to be  
\be
\label{yb}
 \vert \Psi_{qq'} \rangle = \frac{1}{\sqrt{2}} \vert 0_1 0'_2 \rangle + \frac{1}{\sqrt{2}} \vert 1_1 1'_2 \rangle 
\ee
We wish to examine the effects of different noisy environments on the maximum possible entanglement contained in the state $\vert \Psi_{qq'} \rangle$. In order to do this, we employ negativity of partial transpose $N(\rho)$ as the suitable measure of entanglement~\cite{neg1,neg2,neg3}. This measure being necessary and sufficient for $2\times 3$ systems~\cite{neg1,neg2,neg3}, the choice of the measure is justified. 
In the following, we consider amplitude noise, phase noise, Generalized amplitude damping noise and depolarizing noise as  models for noisy environments acting on the state $\vert \Psi_{qq'} \rangle$.  
\subsection{Amplitude damping:} Amplitude damping noise gives the right description for energy dissipation
from a quantum system. It characterizes processes such as spontaneous emission of a photon from a quantum system, attainment of thermal equilibrium by a spin system at a high temperature and is one of the well-studied noise models in the literature~\cite{nc,esd1,esd19}.   
The Kraus operators for a single qubit\cite{esd1}
amplitude noise are given by 
\begin{equation}
E_{0}=\ba{cc}
\eta & 0  \\  0 &  1 \ea
;
E_{1}=
\ba{cc}
0 & 0  \\  \sqrt{1-\eta^2} &  0
\ea
\end{equation}
where $\eta = e^{-\frac{\Gamma t}{2}}$, with $\Gamma$ being the decay factor of the amplitude noise. 
The local action of the amplitude damping channel on the qubit (first subsystem) of the pure maximally entangled state in Eq. (\ref{yb}) can be written as
\begin{equation}
\label{yv}
\rho^{A}_{qq'} = \sum_{i=0}^{1} {(E_{i} \otimes I_3) \rho_{qq'} (E_{i} \otimes I)^{\dagger}}
\end{equation}
where $I_3$ is the $3\times 3$ identity matrix and $ \rho_{qq'} = \vert \Psi_{qq'} \rangle \langle \Psi_{qq'} \vert $ is the density matrix of the state $\vert\Psi_{qq'}\rangle$. 
Explicitly, the state $\rho^{A}_{qq'}$ is given by 
\be
\label{a23}
\rho^{A}_{qq'}=\frac{1}{2}\ba{cccccc}\eta^2 & 0 & 0 & 0 & \eta & 0 \\
0 & 0 & 0 & 0 & 0 & 0 \\
0 & 0 & 0 & 0 & 0 & 0 \\
0 & 0 & 0 & 1-\eta^2 & 0 & 0 \\
\eta & 0 & 0 & 0 & 1 & 0 \\ 
0 & 0 & 0 & 0 & 0 & 0 
\ea.
\ee
The partial transpose $\bar{\rho}^{A}_{qq'}$ obtained by transposing the qubit indices is given by 
\be
\label{a23pt}
\bar{\rho}^{A}_{qq'}=\frac{1}{2}\ba{cccccc}\eta^2 & 0 & 0 & 0 & 0 & 0 \\
0 & 0 & 0 & \eta & 0 & 0 \\
0 & 0 & 0 & 0 & 0 & 0 \\
0 & \eta & 0 & 1-\eta^2 & 0 & 0 \\
0 & 0 & 0 & 0 & 1 & 0 \\ 
0 & 0 & 0 & 0 & 0 & 0 
\ea
\ee
and its non-zero eigenvalues are readily evaluated to be 
\be
\lambda_1=\lambda_2=\frac{1}{2};\ \lambda_3=\frac{\eta^2}{2};\ \lambda_4=-\frac{\eta^2}{2}.
\ee
The only negative eigenvalue of $\bar{\rho}^{A}_{qq'}$ being $\lambda_4$, the negativity of partial transpose is given by 
\begin{equation}
\label{nta}
N(\rho^{A}_{qq'}) = \frac {\eta^2}{2}=\frac{e^{-\Gamma t}}{2} 
\end{equation}
The asymptotic decay of $N(\rho^{A}_{qq'})$ with time is evident through Eq. (\ref{nta}).
It can be recalled here that similar situation arose in the case of the local action of amplitude noise on a $2$-qubit pure state~\cite{esd19}.  
\subsection{Phase damping:}  
The noise that describes the process of loss of quantum information without loss of energy is referred to as phase damping noise~\cite{nc} or dephasing noise. 
It is a uniquely quantum mechanical noise~\cite{nc} and the single qubit Kraus operators of phase damping channel are given by~~\cite{n10} 
\begin{equation}
P_{0}=
\begin{pmatrix}
1 & 0  \\  0 &  \gamma
\end{pmatrix};
P_{1}=
\ba{cc}
0 & 0  \\  0 &  \sqrt{1-\gamma^2}
\ea
\end{equation} 
where $\gamma = e^{\frac{- \Gamma t}{2} }$ and $ \Gamma $ is the decay constant of the dephasing noise.
The time evolved density matrix of $\Psi_{qq'}$ with its first qubit being subjected to phase damping is given by
\begin{equation}
\label{rhoph}
\rho^P_{qq'}=\frac{1}{2}
\ba{cccccc}
1 & 0 & 0 & 0 & \gamma & 0 \\ 0 & 0 & 0 & 0 & 0 & 0 \\ 0 & 0 & 0 & 0 & 0 & 0 \\ 0 & 0 & 0 & 0 & 0 & 0 \\ \gamma & 0 & 0 & 0 & 1 & 0 \\ 0 & 0 & 0 & 0 & 0 & 0
\ea
\end{equation}
The non-zero eigenvalues of the partially transposed density matrix of $\rho^P_{qq'}$, transposed with respect to the qubit indices, are identified to be 
$1/2,\,1/2,\,\gamma/2,\,-\gamma/2$ and hence the negativity of partial tranpose $N(\rho^P_{qq'})$ is given by 
\begin{equation}
\label{ntph}
N(\rho^P_{qq'})= \frac{\gamma}{2}=\frac{1}{2} \exp[-\Gamma t/2]
\end{equation}
The exponential decay of $N(\rho^P_{qq'})$ due to the local action of dephasing noise on the first subsystem (qubit) of the pure maximally entangled qubit-qutrit state $\vert \Psi_{qq'}\rangle$ is thus evident through Eq. (\ref{ntph}). Here too we can recall that the dephasing noise acting on the first (or second) qubit of the pure $2$-qubit state resulted in an asymptotic decay as detailed in Ref.~\cite{esd19}.  
\subsection{Generalized Amplitude Damping :}  Generalized Amplitude Dampling(GAD) is a quantum noise that describes the effect of dissipation to environment at finite temperature~\cite{nc}.  The relaxation processes due to coupling of spins to their surrounding lattice which is at a high temperature than the temperature of the spins are ably modelled by GAD~\cite{nc}. Such processes often occur in the NMR implementation of quantum computers and we believe that the study of the effect of GAD on pure maximally entangled states is also important as regards any future protocols using entangled pure states.   

The Kraus operators corresponding to Generalized Amplitude Damping on a single qubit are represented by~\cite{esdpr5}
\begin{eqnarray}
G_{0}= \sqrt{1-p}
\ba{cc}
1 & 0  \\  0 & \eta
\ea & & 
G_{1}= \sqrt{1-p}
\ba{cc}
0 & \sqrt{1-\eta^2}  \\  0 &  0
\ea \nonumber \\
G_{2}= \sqrt{p}
\ba{cc}
\eta & 0  \\  0 &  1
\ea & & 
G_{3}= \sqrt{p}
\ba{cc}
0 & 0  \\  \sqrt{1-\eta^2} &  0
\ea
\end{eqnarray}
where $\eta = e^{- \Gamma t/2}$ and $ \Gamma $ is the decay factor corresponding to generalized amplitude damping. 
The state $\rho^G_{qq'}  =\sum_{i=0}^{3} {(G_{i} \otimes I_3) \rho_{qq'} (G_{i} \otimes I_3)^{\dagger}}$ and its partial transpose $\bar{\rho}^G_{qq'}$  (transposed either with respect to the qubit or with the qutrit) can be readily evaluated. They are given by 
\begin{small}
\be
\label{g23}
\rho^{G}_{qq'}=\ba{cccccc}\frac{1-p(1-\eta^2)}{2} & 0 & 0 & 0 & \frac{\eta}{2} & 0 \\
0 & \frac{(1-p)(1-\eta^2)}{2} & 0 & 0 & 0 & 0 \\
0 & 0 & 0 & 0 & 0 & 0 \\
0 & 0 & 0 & \frac{p(1-\eta^2)}{2}  & 0 & 0 \\
\frac{\eta}{2} & 0 & 0 & 0 &\frac{p(1-\eta^2)+\eta^2}{2}  & 0 \\ 
0 & 0 & 0 & 0 & 0 & 0 
\ea \nonumber 
\ee
\be
\label{g23pt}
\bar{\rho}^{G}_{qq'}=\ba{cccccc}\frac{1-p(1-\eta^2)}{2} & 0 & 0 & 0 & 0 & 0 \\
0 & \frac{(1-p)(1-\eta^2)}{2} & 0 & \frac{\eta}{2} & 0 & 0 \\
0 & 0 & 0 & 0 & 0 & 0 \\
0 & \frac{\eta}{2} & 0 & \frac{p(1-\eta^2)}{2}  & 0 & 0 \\
\frac{\eta}{2} & 0 & 0 & 0 &\frac{p(1-\eta^2)+\eta^2}{2}  & 0 \\ 
0 & 0 & 0 & 0 & 0 & 0 
\ea \nonumber 
\ee 
\end{small} 

An explicit evaluation of $N(\rho^{G}_{qq'})$ shows that an asymptotic decay of entanglement happens at $p=0$, $p=1$ and this is expected because the GAD reduces to amplitude damping when $p=0$ or $1$. Also, the variation of $N(\rho^{G}_{qq'})$ with $\Gamma t$ can be seen to be symmetric over the value $p=1/2$.
A plot of the negativity of partial transpose $N(\rho^{G}_{qq'})$ versus $\Gamma t$ for different values of $p$ are as shown in Fig. 1. 
\begin{figure}[ht] 
\includegraphics*[width=2.2in,keepaspectratio]{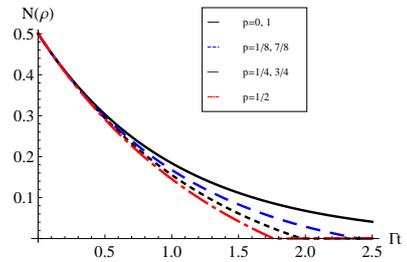} 
\caption{Variation of  $N(\rho^{G}_{qq'})$ with respect to $\Gamma t$ for different values of $p$. The sudden death of entanglement is seen to occur when $0<p<1$ and the decay time is shortest when $p=1/2$.}
\end{figure}  

It is readily seen through Fig.1 that there is  finite time disentanglement due to the local action of GAD on the qubit of maximally entangled pure qubit-qutrit state $\vert \Psi_{qq'}\rangle$ when $0<p<1$.  It can also be seen that the decay time is shortest when $p=1/2$.  We therefore conclude that GAD is capable of causing sudden death of entanglement in pure maximally entangled qubit-qutrit states due to its action on one of the subsystems.  
 
\subsection{Depolarizing Noise:} A quantum noise that is capable of {\emph{depolarizing}} a qubit with probability $\alpha$ and keeping it undisturbed with a probability $1-\alpha$ is the depolarizing noise~\cite{nc}. Here `depolarizing' a qubit means converting the qubit into a completely mixed state $I_2/2$~\cite{nc}. 

The single qubit Kraus operators~\cite{esd14} for depolarizing noise are given by,
\begin{eqnarray}
D_{0}= \sqrt{1-\alpha}
\ba{cc}
1 & 0  \\  0 &  1
\ea & & 
D_{1}=\sqrt{\frac{\alpha}{3}}
\ba{cc}
0 & 1  \\  1 &  0
\ea \nonumber \\
D_{2}= \sqrt{\frac{\alpha}{3}}
\ba{cc}
0 & -i  \\  i & 0
\ea & &
D_{3}= \sqrt{\frac{\alpha}{3}}
\ba{cc}
1 & 0  \\  0 &  -1
\ea;
\end{eqnarray}
Here $\alpha= 1-e^{- \Gamma t/2}$ and $ \Gamma $ is the  decay factor of the depolarizing noise.
On evaluating $\rho^D_{qq'}=\sum_{i=1}^{4} {(D_{i} \otimes I_3) \rho_{qq'} (D_{i} \otimes I_3)^{\dagger}}$ and its partial transpose $\bar{\rho}^D_{qq'}$, we readily obtain the negativity of partial transpose of $\rho^D_{qq'}$ as
\be
N(\rho^D_{qq'}) = \frac{(1- 2\alpha)}{2}=\frac{2\exp[-\Gamma t/2]-1}{2}.
\ee
Notice that $N(\rho^D_{qq'})$ becomes zero when $ \alpha=\frac{1}{2} $ or when $ \Gamma t = 1.3863$. Thus, we can conclude that depolarizing noise causes finite time disentanglement in the pure maximally entangled state under consideration, even when acting only on one of the subsystems of the state. 

As both GAD and depolarizing noise induce entanglement sudden death in the state $\vert \Psi_{qq'}\rangle$, we wish to examine which among these is more effective. The following graph (Fig. 2) compares the entanglement decay time of GAD (for $p=1/2$) and depolarizing noise. 
\begin{figure}[ht]
\includegraphics*[width=2.2in,keepaspectratio]{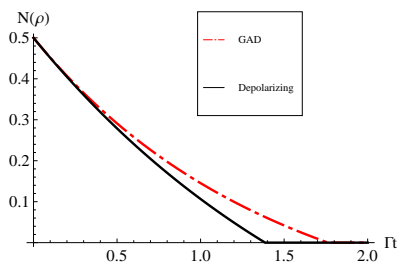} 
\caption{The negativity of partial transpose of maximally entangled pure qubit-qutrit state $\vert \Psi_{qq'}\rangle$ with its first qubit being under the action of generalized amplitude damping and depolarizing noise are plotted as a function of time. The sudden death of entanglement occurs faster in the case of depolarizing noise than in the case of GAD.}
\end{figure} 

Fig. 2 clearly indicates that depolarizing causes an early onset of entanglement sudden death than that due to Generalized Amplitude Damping noise when acting on the qubit of pure maximally entangled state qubit-qutrit state. 

While we have considered the action of different noise models on the first subsystem (qubit) of the qubit-qutrit state $\vert\Psi_{qq'}\rangle$, it is not difficult to see that the results will not be any different when the second subsystem (qutrit) alone is exposed to noise. In the following section, we examine whether the results obtained here in the case of $2\times 3$ dimensional pure maximally entangled states hold good  even for higher dimensional states and carry out an analysis on the effects of different noises on pure maximally entangled $3\times 3$ dimensional states. 

\section{Local action of noise on pure qutrit-qutrit states} 
A maximally entangled pure qutrit-qutrit state is given by, 
\be 
\label{yu}
\vert \Phi \rangle_{q'q'} = \frac{1}{\sqrt{3}} ( \vert 0'_1 0'_2 \rangle +  \vert 1'_1 1'_2 \rangle +\vert 2'_1 2_2' \rangle).
\ee
with $\vert i'_r\rangle$, $i=0',\,1',\,2'$, $r=1,\,2$ being the basis vectors of the $3$-dimensional qutrit space.  
The Kraus operators of amplitude damping for a qutrit~\cite{amp3O,amp3} are given by
\begin{eqnarray}
E_{0}&=&
\ba{ccc}
1 & 0 & 0\\  0 &\eta & 0 \\ 0 & 0 & \eta
\ea, \  
E_{1}=
\ba{ccc}
0 & \sqrt{1-\eta^2} & 0\\  0 & 0 & 0 \\ 0 & 0 &  0
\ea \nonumber \\
E_{2}&=&
\ba{ccc}
0 & 0 & \sqrt{1-\eta^2} \\  0 & 0 & 0 \\ 0 & 0 &  0
\ea; \ \ \eta = e^{\frac{-\Gamma t}{2}}.
\end{eqnarray}
On explicit evaluation, the negativity of partial transpose of the state $\rho^{A}_{q'q'} = \sum_{i=0}^{2} {(E_{i} \otimes I_3) \rho_{q'q'} (E_{i} \otimes I_3)^{\dagger}}$, where  $\rho_{q'q'}=\vert \Phi_{q'q'}\rangle \langle \Phi_{q'q'}\vert$, is obtained as   
\begin{equation}
N(\rho^{A}_{q'q'})=\frac{5}{6} \eta^2=\frac{5}{6} e^{-\Gamma t}
\end{equation}
This indicates asymptotic decay of entanglement for the pure qutrit-qutrit state $\rho^{A}_{q'q'}$ as in the case of qubit-qutrit state $\rho^{A}_{qq'}$ subjected to local action of amplitude noise. 

The action of phase noise on a single qutrit can be represented by the Kraus operators
\begin{eqnarray}
P_{0}&=&
\ba{ccc}
1 & 0 & 0\\  0 & \gamma & 0 \\ 0 & 0 &  \gamma
\ea,\ 
P_{1}=
\ba{ccc}
0 & 0 & 0\\  0 & \sqrt{1-\gamma^2} & 0 \\ 0 & 0 &  0
\ea \nonumber \\ 
P_{2}&=&
\ba{ccc}
0 & 0 & 0 \\  0 & 0 & 0 \\ 0 & 0 &  \sqrt{1-\gamma^2}
\ea; \ \ \gamma=e^{\frac{- \Gamma t}{2} }
\end{eqnarray}
Here too, as in the case of amplitude noise, the negativity of partial transpose given by
\be
N(\rho^{P}_{q'q'})=\frac{\gamma}{3}(\gamma+2)=\frac{e^{-\Gamma t/2}}{3}\left(e^{-\Gamma t/2}+2 \right)
\ee
shows asymptotic decay of entanglement.

The single qutrit Kraus operators for Generalized Amplitude Damping noise are given by,
\begin{eqnarray}
G_{0}&=&\sqrt{p}
\ba{ccc}
1 & 0 & 0 \\  0 &  \eta & 0 \\ 0 & 0 &  \eta
\ea, \ 
G_{1}= \sqrt{p}
\ba{ccc}
0 & \sqrt{1-\eta^2} & 0  \\  0 & 0 & 0 \\ 0 & 0 & 0
\ea \nonumber \\
G_{2}&=& \sqrt{p}
\ba{ccc}
0 & 0 & \sqrt{1-\eta^2}  \\  0 & 0 & 0 \\ 0 & 0 & 0
\ea,\ 
G_{3}= \sqrt{1-p}
\ba{ccc}
\eta & 0 & 0  \\  0 & \eta & 0 \\ 0 &  0 & 1 
\ea \nonumber \\
G_{4}&=& \sqrt{1-p}
\ba{ccc}
0 & 0 & 0  \\ 0 & 0 & 0  \\  0 & \sqrt{1-\eta^2} & 0
\ea,  \\ 
G_{5}&=& \sqrt{1-p}
\ba{ccc}
0 & 0 & 0  \\ 0 & 0 & 0  \\  \sqrt{1-\eta^2} & 0 & 0
\ea   \nonumber  
\end{eqnarray}
with $ \eta=e^{\frac{- \Gamma t}{2}}$. 
A plot of the negativity of partial transpose $N(\rho^{G}_{q'q'})$ versus $\Gamma t$ for different values of $p$ are as shown in Fig. 3. 
\begin{figure}[ht] 
\includegraphics*[width=2.2in,keepaspectratio]{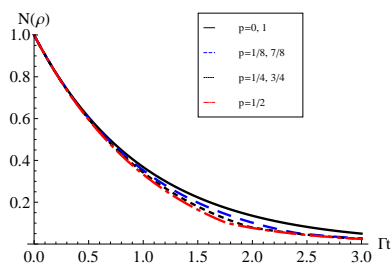} 
\caption{Variation of  $N(\rho^{G}_{q'q'})$ with respect to dimensioless time parameter $\Gamma t$ for different values of $p$.}
\end{figure}  
 
It can be seen that under the action of GAD on a single qutrit of the state $\vert \Phi_{q'q'} \rangle$, the decay of entanglement is almost smooth (See Fig. 3). 
The variation is symmetric about $p=1/2$ and as the GAD reduces to amplitude damping noise when $p=0$, $p=1$, an exponential decay occurs for both these values.   
   
The single qutrit Kraus operators for depolarizing noise~\cite{esd14,dep2} are given by,
\begin{eqnarray}
D_{0} &=& \sqrt{1-\alpha}\, I_3,\ \  D_{1}= \sqrt{\frac{\alpha}{8}}\, Y,  \nonumber \\ 
D_{2}&=& \sqrt{\frac{\alpha}{8}}\, Z,\ \  D_{3} =\sqrt{\frac{\alpha}{8}}\, Y^2, \nonumber \\
D_{4}&=&\sqrt{\frac{\alpha}{8}}\, YZ,\ \  D_{5}= \sqrt{\frac{\alpha}{8}}\, Y^2 Z, \\
D_{6} &=& \sqrt{\frac{\alpha}{8}}\, Y Z^{2},\ \  D_{7}=\sqrt{\frac{\alpha}{8}}\, Y^{2}Z^{2} \nonumber \\
D_{8}& =&\sqrt{\frac{\alpha}{8}}\, Z^{2} \nonumber 
\end{eqnarray}
where $I_3$ is the $3\times 3$ Identity matrix and 
\begin{eqnarray}
Y &=& \ba{ccc}
0 & 1 & 0 \\ 0 & 0 & 1 \\ 1 & 0 & 0 \ea,\      
Z = \ba{ccc}
1 & 0 & 0 \\ 0 & \omega & 0 \\ 0 & 0 & \omega^2 \ea; \\
& & \omega^3=1; \ \ \alpha= 1-e^{\frac{- \Gamma t}{2}}. \nonumber 
\end{eqnarray}
The negativity of partial transpose of the state $\rho^{D}_{q'q'} = \sum_{i=0}^{2} {(D_{i} \otimes I_3) \rho_{q'q'} (D_{i} \otimes I_3)^{\dagger}}$ is obtained as 
\begin{equation*}
N(\rho^{D}_{q'q'}) = \frac{1}{2}(2-3\alpha)=\frac{1}{2}(3 e^{-\Gamma t/2}-1)
\end{equation*}
The sudden death of entanglement caused by depolarizing noise is evident, as $N(\rho^{D}_{q'q'})=0$  when $ \alpha = \frac{2}{3} $ or when $ \Gamma t = 2.2$.  
In order to compare the entanglement decay in the case of GAD and depolarizing noise, we have plotted below (Fig. 4) the negativity of partial transpose in both cases with respect to time.  
\begin{figure}[ht]
\includegraphics*[width=2.2in,keepaspectratio]{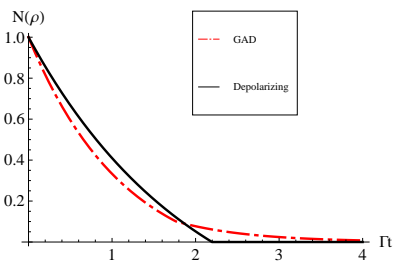} 
\caption{The variation of negativity of partial transpose with respect to dimensionless time parameter $\Gamma t$  under the local action of depolarizing and generalized amplitude damping.}
\end{figure} 

The fact that depolarizing noise is more effective than GAD in disentangling a pure maximally entangled qutrit-qutrit state, even when acting on a single qutrit of the state, is clearly shown in Fig. 4. 

It is to be mentioned here that though negativity of partial transpose is only a sufficient criterion for entanglement of $3\times 3$ dimensional states, the analysis done in this section is a definite indication of the effect of local action of different noise models on one of the subsystems of a pure qutrit-qutrit state.   
\section{Conclusion} 

In this article, we have analyzed the effects of various noisy environments on a pure maximally entangled $2\times 2$ and $3\times 3$ systems when only one subsystem is exposed to noise.  We have shown that amplitude noise and dephasing noise can only cause asymptotic decay of entanglement whereas 
both generalized amplitude damping and depolarizing noise induce finite time disentanglement in the maximally entangled pure states. A comparison of timescales of disentnaglement reveals that depolarizing causes sudden death of entanglement faster than generalized amplitude damping. We have thus proved the effectiveness of  depolarizing noise, over other noisy environments, in causing entanglement sudden death in pure maximally entangled states even while acting on only part of the system. Such an observation was done in Ref.~\cite{esd19} by examining the action of different noises individually on a part of $2$ qubit pure states.  
This work also strengthens the assertion~\cite{esd19} that composite noise is not necessary in causing sudden death of entanglement by showing that higher dimensional pure states also get disentangled in finite time due to the action of a chosen noise on a part of the state. Our choice of the pure states in our analysis is to rule out any prior interactions with the environment that might have contributed to the finite time disentanglement~\cite{esd19}. Thus, the sudden death of entanglement caused by depolarizing noise (or GAD) is due to that noise alone and not due to any previous interaction with the environment. We believe that this analysis is helpful in finding out proper means to avoid disentanglement due to destructive noisy environment such as depolarizing noise.    
\begin{center}
{\bf Acknowledgment}
\end{center}
K.O. Yashodamma and P.J. Geetha acknowledge the support of Department of Science and
Technology (DST), Govt. of India through the award
of INSPIRE fellowship.

\end{document}